# CaraNet: Context Axial Reverse Attention Network for Segmentation of Small Medical Objects

Ange Lou[1], Shuyue Guan[2], Murray Loew[2]

Vanderbilt University[1], Nashville TN, USA

ange.lou@vanderbilt.edu

The George Washington University[2], Washington DC, USA

{frankshuyueguan, loew}@gwu.edu

**Abstract**

**Purpose**: Segmenting medical images accurately and reliably is important for disease diagnosis and treatment. It is a challenging task because of the wide variety of objects' sizes, shapes, and scanning modalities. Recently, many convolutional neural networks (CNN) have been designed for segmentation tasks and achieved great success. Few studies, however, have fully considered the sizes of objects, and thus most demonstrate poor performance for small objects segmentation. This can have a significant impact on the early detection of diseases.

**Approach**: This paper proposes a Context Axial Reverse Attention Network (CaraNet) to improve the segmentation performance on small objects compared with several recent state-of-the-art models. CaraNet applies axial reserve attention (ARA) and channel-wise feature pyramid (CFP) module to dig feature information of small medical object. And we evaluate our model by six different measurement metrics.

**Results**: We test our CaraNet on brain tumor (BraTS 2018) and polyp (Kvasir-SEG, CVC-ColonDB, CVC-ClinicDB, CVC-300, and ETIS-LaribPolypDB) segmentation datasets. Our CaraNet achieves the top-rank mean Dice segmentation accuracy, and results show a distinct advantage of CaraNet in the segmentation of small medical objects.

**Conclusion**: We proposed CaraNet to segment small medical objects and outperform other state-of-the-art methods.

Codes available: https://github.com/AngeLouCN/CaraNet

## 1. INTRODUCTION

Deep learning has had a tremendous impact on various fields in science. Our focus of the current study in deep learning is on one of the most critical areas of computer vision: medical image segmentation. Recently, various convolutional neural networks (CNNs) have shown great performance on medical image segmentation [1,2,3,4]. Those CNNs have been introduced for various medical imaging modalities, including X-ray, visible-light imaging, magnetic resonance imaging (MRI), positron emission tomography (PET), and computerized tomography (CT). They all achieved excellent performance on medical image segmentation challenges from different modalities, like BraTS [5,6,7], KiTS19 [8], and COVID19-20 [9,10]. To obtain more accurate segmentation results, many works introduced improvements of network architectures. Those improvements are mostly attributed to exploring new neural architectures by designing networks with varying depths (ResNet [11]), widths (ResNeXt [12]), connectivity (DenseNet [13] and GoogLeNet [14]), or new types of components (pyramid scene [15] and atrous convolution [16]). Although those new architectures improve the overall segmentation results, they are less sensitive to detecting small medical objects. And it is very common in medical image segmentation that the anatomy of interest occupies only a very small portion of the image [17]. Most extracted features belong to the background, while these small lesion areas are important for early detection and diagnosis. For example, the survival rate decreases with the growing size of a brain tumor [18]. Thus, it has clinical significance to build an effective network to detect tiny medical objects.

The attention mechanism plays a dominant role in neural network research. It can effectively use information transferred from several subsequent feature maps to detect the salience features [19]. Many attention methods such as self-attention and multi-head attention have been verified to have high performance in applications of natural language processing [20] and computer vision [21]. Those attention methods also have been successfully used for medical image segmentation; for example, the Medical Transformer [22] (MedT) used a gated axial self-attention layer to build a Local-Global (LoGo) network for ultrasound and microscopy image segmentation, TransUNet [23] stacked self-attention as a transformer in the encoder for CT image segmentation, and CoTr [24] bridged two CNN encoder and decoder by the transformer encoder for multi-organ segmentation. All those attention-based segmentations achieve significant improvement compared with purely convolutional neural networks like U-Net [1] and FCN [25].

Although those new types of neural networks show good performance on many medical segmentation tasks, they seldom consider the small object segmentation, especially in the medical image area. We propose here a novel attention-based deep neural network, called **C**ontext **A**xial **R**everse **A**ttention **Net**work (**CaraNet**). The contribution of the paper can be summarized as follows:
1) We propose a novel neural network – CaraNet -- to solve the problem of segmentation of small medical objects.
2) We introduce a method to evaluate the network's performance on small medical objects.
3) Our experiments show that CaraNet outperforms most current models (e.g., DS-TransUNet from TMI '22, CCBANet from MICCAI '21and PraNet from MICCAI '20) and advances the state-of-the-art by a large margin, both overall and on small objects, in segmentation performance on polyps.

## 2. METHOD

Figure 1 shows the architecture of our CaraNet, which uses a parallel partial decoder [26] to generate the high-level semantic global map and a set of context and axial reverse attention operations to detect global and local feature information. We introduce main components of the CaraNet architecture in the following subsections with explanation of the motivation, purpose, or effectiveness to add these components.

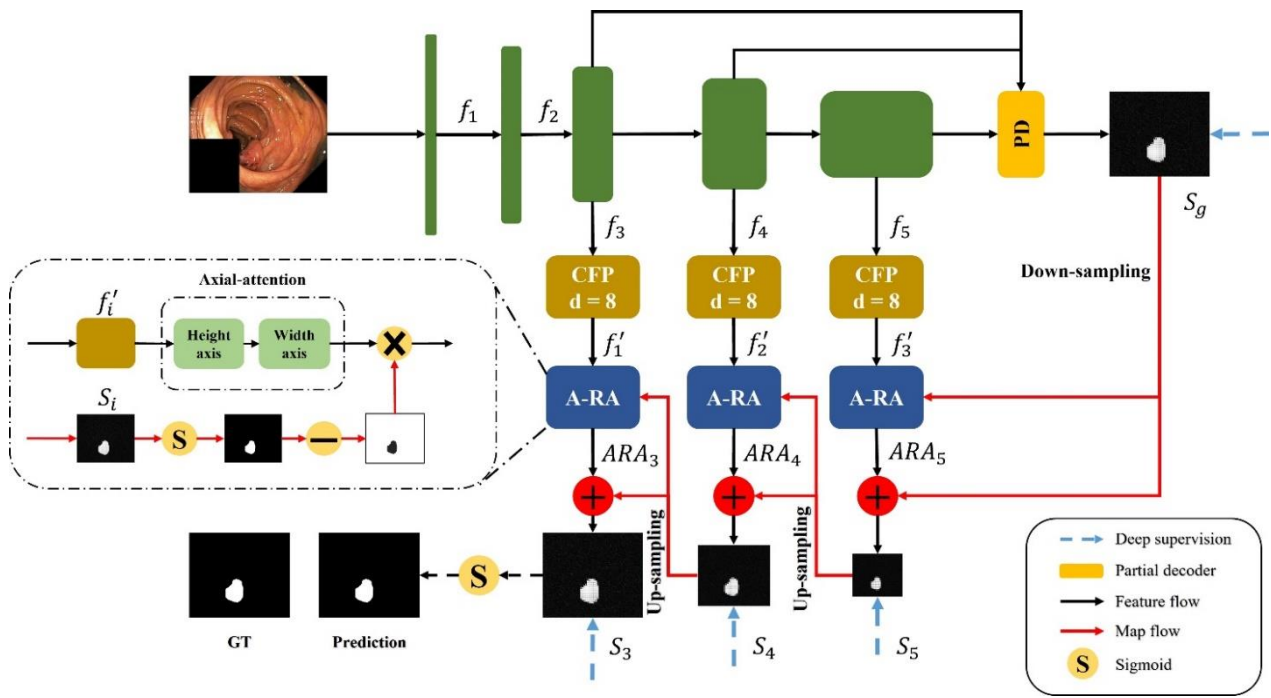

Figure 1. Overview of CaraNet, which contains pretrained backbone, partial decoder, channel-wise feature pyramid (CFP) module and axial reverse attention (A-RA) module.

### 2.1 Backbone

Transfer learning provides a feasible method for alleviating the challenge of data-hunger, and it has been widely applied to the field of computer vision [27]. Benefiting from the strong visual knowledge distributed in ImageNet [28], the pre-trained CNNs can be fine-tuned with a small amount of task-specific data and can perform well on downstream tasks. Since Res2Net [29] can construct hierarchical residual-like connections within one single residual block that has stronger multi-scale representation ability, we applied the pre-trained Res2Net as the backbone of CaraNet.

**2.2 Partial decoder**

Existing state-of-the-art segmentation networks rely on aggregating multi-level features from the encoder (e.g., U-Net aggregates all level features extracted from an encoder). Compared to the high-level features, however, low-level features contribute less to performance but have higher computational cost because of their larger spatial resolution [30]. Thus, we applied the parallel partial decoder [26] as shown in Figure 2 to aggregate high-level features. We feed the original image which size is $h \times w \times c$ ($h$, $w$, and $c$ represent the height, width, and channel) into Res2Net, and we can get five different level features $\{f_i, i = 1, ..., 5\}$ with resolution $\{\frac{h}{2^{i-1}}, \frac{w}{2^{i-1}}\}$. We aggregated the high-level features $\{f_3, f_4, f_5\}$ from Res2Net by using the partial decoder with a parallel connection. Then, we can get a global map $S_g = PD(f_3, f_4, f_5)$.

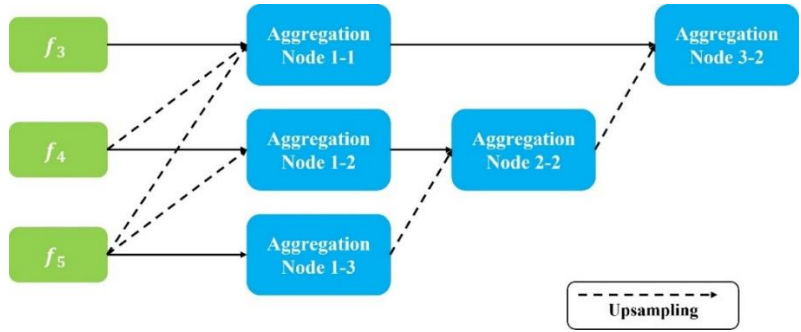

Figure 2. Overview of partial decoder with parallel connection.

**2.3 Channel-wise feature pyramid module**

The Feature Pyramid (FP) has been widely used in deep learning models for computer vision tasks due to its ability to represent multi-scale features. For example, PSPNet [31] builds a pyramid pooling module with different sizes' pooling layers to extract multi-scale features, and the Feature Pyramid Network (FPN) [32] takes different strides with convolution kernels to obtain a FP. Although those FP-based methods perform well in the computer vision area, they cannot avoid using large numbers of parameters, which consume a large amount of computation resources. In addition, their receptive fields are usually small and do not perform well in datasets with sharply varying object sizes [33]. Alternatively, our previous works [33, 34] proposed a lightweight Channel-wise Feature Pyramid (CFP) module and successfully applied it to both nature and medical image segmentation. The architecture of this CFP module is shown in Figure 3.

Figure 3(a) shows the architecture of the CFP module; it contains total $K$ channels and each channel has its own dilation rate $r_K$. Typically, we choose the $K = 4$ for CaraNet and the dilation rates for each channel $\{r_1, r_2, r_3, r_4\} = \{1,2,4,8\}$ which has been verified as the best dilation rates combination for CaraNet in Table 4; thus, each channel's dimension is $M/4$. Simple feature fusion method sometimes introduces some unwanted checkboard or gridding artifacts that greatly influence the accuracy and quality of segmentation masks [33]. Thus, we applied hierarchical feature fusion [35] (HFF) to sum the outputs of all channels step by step. For the FP channel, we provide two versions with regular convolution and asymmetric convolution as shown in Figure 3(b) and (c). We connected the outputs of each convolutional module by using skip connection, and thus each channel can be considered as a sub-pyramid. We selected the regular convolution as FP channel for CaraNet. The overall FP is obtained from concatenating those sub-pyramids from the hierarchical feature fusion operation. The final FP contains four levels of feature stacks as shown in Figure 4. These four levels of feature stacks $\{level_i, i = 1, ..., 4\}$ are computed by:

$$\begin{cases} level_1 = out_{FP1} \\ level_2 = level_1 + out_{FP2} \\ level_3 = level_2 + out_{FP3} \\ level_4 = level_3 + out_{FP4} \end{cases} \quad (1)$$

And the final FP is computed by $\sum_i level_i$. Based on our split-merge feature pyramid strategy, the receptive fields of a single CFP module with dilation rates $\{r_1, r_2, r_3, r_4\} = \{1,2,4,8\}$ varies from $3 \times 3$ to $55 \times 55$, which successfully overcomes the challenge from sharply varying object sizes.

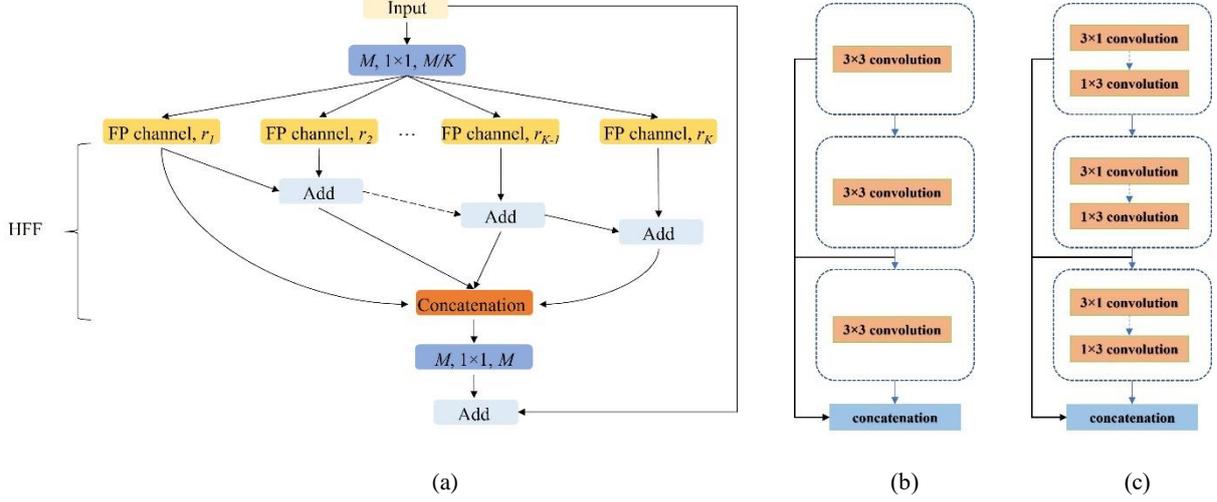

Figure 3. (a) CFP module, (b) FP channel with regular convolution, (c) FP channel with asymmetric convolution

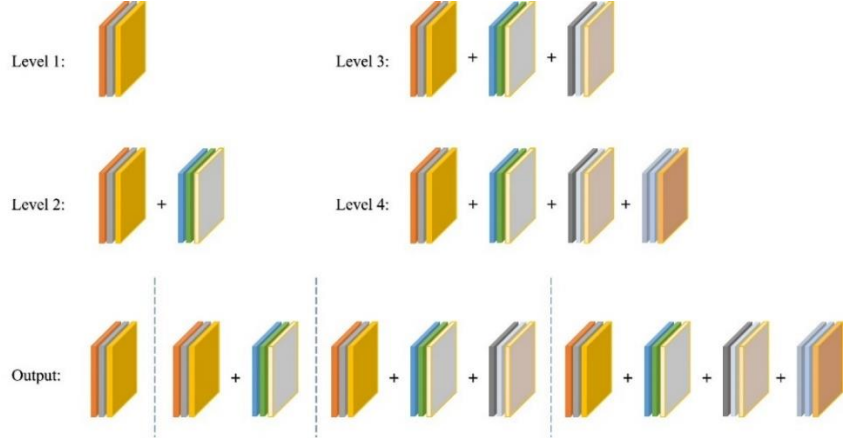

Figure 4. Final feature pyramid obtained from CFP module.

**2.4 Axial reverse attention module**

The previous partial decoder that generates the global map $S_g$ (Sec. 2.2) could roughly locate the position of medical objects, and the CFP module extracted only multi-scale features from the pre-trained model. To obtain more accurate feature information, we designed the Axial Reverse Attention (A-RA) module to refine localization information and multi-scale features efficiently. The overview and detail of the A-RA module can be seen in Figure 1 and Figure 5, respectively. The input of the top line is the multi-scale feature maps $f_i'$ from the CFP module and we used axial attention to analyze the salience information. The axial attention is based on self-attention, which maps a query and a set of key-value pairs to an output and the operation:

$$Attention(Q, K, V) = Softmax\left(\frac{QK^T}{\sqrt{d_K}}\right) \qquad (2)$$

where $Q, K, V$, and $d_K$ represent query, key, value, and dimension of key, respectively. However, self-attention consumes great computational resources, especially when the spatial dimension of the input is large [36]. Therefore, we applied axial attention, which factorizes 2D attention into two 1D attention along height and width axes. Here we replace the softmax activation function with a sigmoid, based on the experiments. For the second line, we applied the reverse operation [37] to detect the salience features from the side-output $S_i$, which is obtained from the output of the previous A-RA module. The reverse operation is:

$$R_i = 1 - Sigmoid(S_i). \qquad (3)$$

The total axial reverse attention operation is:

$$ARA_i = AA_i \odot R_i \qquad (4)$$

where $\odot$ is element-wise multiplication, and the $AA_i$ is feature from the axial attention route.

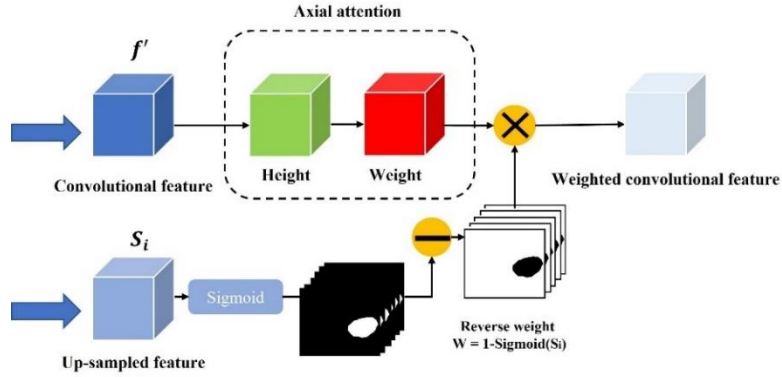

Figure 5. Structure of Axial Reverse Attention module.

**2.5 Deep supervision**

We apply weighted intersection over union (IoU) and weighted binary cross-entropy (BCE) in our loss function: $\mathcal{L} = \mathcal{L}_{IoU}^w + \mathcal{L}_{BCE}^w$ to calculate the global loss and local (pixel-level) loss, respectively. To train CaraNet, we apply deep supervision for the three side-outputs $(S_1, S_2, S_3)$ and the global map $S_g$. Before calculating the loss, we up-sampled them to the same size as ground truth $G$. Thus, the total loss:

$$\mathcal{L}_{total} = \mathcal{L}\left(G, S_g^{up}\right) + \sum_{i=3}^{5} \mathcal{L}\left(G, S_i^{up}\right) \qquad (5)$$

**2.6 Small object segmentation analysis**

Since the size of all images input to the segmentation models must be fixed, the size of an object is determined by the number of pixels in the object $m$ and the number of total pixels in the image $N$. Thus, we consider the object's size using the size ratio (proportion) = $m/N$. Then, we evaluate the performance of segmentation models according to the sizes of objects. Especially, we mainly focus on the small areas whose size ratios are smaller than 5%. To define the watershed of small size is a question; it depends on data and model performance. We further discuss this question in Discussion section.

To evaluate the performance of segmentation models according to the sizes of objects, we first obtain the mean-Dice coefficients and size ratios of segmentations from the test dataset. Similar to computing the histogram, we plot the results in a curve whose y-axis is mean-Dice coefficients and x-axis is increasingly sorted size ratios. To smooth the curve, we take interval-averaged mean-Dice coefficients by sorted size ratios: we divide the entire range of size ratios into a consecutive, non-overlapping, and of equal length series of intervals, and then calculate the average mean-Dice coefficients of size ratios in each interval. The interval-averaged coefficients have a smooth curve and are more stable in the presence of noise.

## 3. EXPERIMENT

### 3.1 Implementation details

We implemented our model in PyTorch accelerated by the NVIDIA RTX 2070Ti GPU. We resized input images to $352 \times 352$ for polyp segmentation and $256 \times 256$ for brain tumor segmentation and employed a multi-scale training strategy $\{0.75, 1.0, 1.25\}$ instead of data augmentation. We used Adam optimizer with the initial learning rate $1e^{-4}$.

### 3.2 Dataset

We test our CaraNet on five polyp segmentation datasets: ETIS [38], CVC-ClinicDB [39], CVC-ColonDB [40], EndoScene [41], and Kvasir [42]. The first four are standard benchmarks, and the last one is the largest dataset, which was released recently. We also test our model on the **Bra**in **T**umor **S**egmentation 2018 (BraTS 2018) dataset [48, 49], which contains more extremely small medical objects. Table 1 shows the details of these datasets: image size, scale of testing set, and size ratios of medical objects.

The data of brain tumor segmentation is from the multimodal brain tumor segmentation challenge 2018 (BraTS 2018) built by the Section for Biomedical Image Analysis (SBIA) at the University of Pennsylvania. It contains the multimodal brain MRI scans and manual ground truth labels of glioblastoma from 285 cases (patients). Each case includes four scan-modals: 1) native (T1), 2) T1 contrast enhanced (T1ce), 3) T2-weighted (T2), and 4) T2 Fluid Attenuated Inversion Recovery (FLAIR). And each case includes three types of ground truth labels: necrotic and non-enhancing tumor core (NET), gadolinium-enhancing tumor (ET), and peritumoral edema (Ed). In this study, **T1ce** is selected as our input images and ground truth labels use **NET** type because it delineates the minimum areas for small object segmentation. The MRI scans for each case are sliced to 2-D images. As shown in Table 1, the test samples are chosen by the sizes of objects in images (by examining the areas of truth labels) ranging in 0.01% - 4.91%.

Table 1. Details of datasets

|  | **Image size** | **Number of test samples** | **Object size ratio** |
|---|---|---|---|
| **ETIS** | $966 \times 1225$ | 196 | 0.11% - 29.05% |
| **CVC-ClinicDB** | $288 \times 384$ | 62 | 0.34 % - 45.88% |
| **CVC-ColonDB** | $500 \times 574$ | 380 | 0.30% - 63.15% |
| **CVC-300** | $500 \times 574$ | 60 | 0.55% - 18.42% |
| **Kvasir** | $1070 \times 1348$ | 100 | 0.79% - 62.13% |
| **BraTS 2018** | $256 \times 256$ | 3231 | 0.01% - 4.91% |

### 3.3 Baseline

We compared CaraNet with six medical image segmentation models, including state-of-the-art models: U-Net [1], U-Net++ [2], ResUNet-mod [43], ResUNet++ [3], SFA [44], PraNet [27], CCBANet [50] and DS-TransUNet [51].

### 3.4 Training and measurement metrics

We randomly split 80% of images from Kvasir and CVC-ClinicDB as training set and the remainder as a testing dataset. In addition to mean Dice and mean IoU, we also apply four other measurement metrics: weighted dice metric $F_\beta^w$, MAE, enhanced alignment metric $E_\phi^{max}$ [45], and structural measurement $S_\alpha$ [46]. Table 2 shows the polyp segmentation on the five datasets. The weighted dice metric $F_\beta^w$ is used to amend the "equal importance flaw" in dice.

The MAE is used to measure the pixel-to-pixel accuracy. The recently released enhanced alignment metric $E_\phi^{max}$ is utilized to evaluate the pixel-level and global-level similarity. And $S_\alpha$ is used to measure the structure similarity between predictions and ground truth.

### 3.5 Results

We first report the polyp segmentation results and compare with state-of-the-arts such as DS-TransUNet, CCBANet and PraNet in Table 2. Among all five public endoscopy segmentation datasets, our proposed CaraNet achieves best performance, especially in the ETIS dataset which contains more small polyps. We also show some polyp segmentation results in Figure 6. For the five polyp datasets, CaraNet not only outperforms the compared models in overall performance, but also on samples with small polyps. Figure 7 shows the segmentation performance of CaraNet and PraNet for small objects (proportions $\leq 5\%$). For the extremely small object segmentation analysis on the BraTS 2018 dataset, we compare only CaraNet with PraNet because PraNet has the closest performance to ours, and the overall accuracies of the other segmentation models are clearly lower than those of CaraNet and PraNet. (Note: the fluctuations with size in colonoscopy datasets are caused by types and boundary of polyps, and quality of imaging)

Table 2. Quantitative results on Kvasir, CVC-ClinicDB, CVC-ColonDB, ETIS, and CVC-T (test dataset of EndoScene). Note: mDice: mean Dice, mIoU: mean IoU, $F_\beta^w$: weighted dice, $S_\alpha$: structural measurement [46], $E_\phi^{max}$: enhanced alignment [45] and MAE: mean absolute error. ↑ denotes higher the better and ↓ denotes lower the better.

| | Methods | mDice ↑ | mIoU ↑ | $F_\beta^w$ ↑ | $S_\alpha$ ↑ | $E_\phi^{max}$ ↑ | MAE ↓ |
|---|---|---|---|---|---|---|---|
| Kvasir | UNet | 0.818 | 0.746 | 0.794 | 0.858 | 0.893 | 0.055 |
| | UNet++ | 0.821 | 0.743 | 0.808 | 0.862 | 0.910 | 0.048 |
| | ResUNet-mod | 0.791 | n/a | n/a | n/a | n/a | n/a |
| | ResUNet++ | 0.813 | 0.793 | n/a | n/a | n/a | n/a |
| | SFA | 0.723 | 0.611 | 0.670 | 0.782 | 0.849 | 0.075 |
| | PraNet | 0.898 | 0.840 | 0.885 | 0.915 | 0.948 | 0.030 |
| | CCBANet | 0.902 | 0.845 | 0.887 | 0.916 | 0.952 | 0.032 |
| | DS-TransUNet | 0.913 | 0.857 | 0.902 | 0.923 | 0.963 | 0.023 |
| | **CaraNet** | **0.918** | **0.865** | **0.909** | **0.929** | **0.968** | **0.023** |
| CVC-ClinicDB | UNet | 0.823 | 0.755 | 0.811 | 0.889 | 0.954 | 0.019 |
| | UNet++ | 0.794 | 0.729 | 0.785 | 0.873 | 0.931 | 0.022 |
| | ResUNet-mod | 0.779 | n/a | n/a | n/a | n/a | n/a |
| | ResUNet++ | 0.796 | 0.796 | n/a | n/a | n/a | n/a |
| | SFA | 0.700 | 0.607 | 0.647 | 0.793 | 0.885 | 0.042 |
| | PraNet | 0.899 | 0.849 | 0.896 | 0.936 | 0.979 | 0.009 |
| | CCBANet | 0.909 | 0.856 | 0.903 | 0.939 | 0.971 | 0.010 |
| | DS-TransUNet | 0.912 | 0.859 | 0.908 | 0.936 | 0.976 | 0.007 |
| | **CaraNet** | **0.936** | **0.887** | **0.931** | **0.954** | **0.991** | **0.007** |
| ColonDB | UNet | 0.512 | 0.444 | 0.498 | 0.712 | 0.776 | 0.061 |
| | UNet++ | 0.483 | 0.410 | 0.467 | 0.691 | 0.760 | 0.064 |
| | SFA | 0.469 | 0.347 | 0.379 | 0.634 | 0.765 | 0.094 |
| | PraNet | 0.709 | 0.640 | 0.696 | 0.819 | 0.869 | 0.045 |
| | CCBANet | 0.758 | 0.675 | 0.736 | 0.842 | 0.880 | 0.042 |
| | DS-TransUNet | 0.762 | 0.682 | **0.738** | 0.829 | 0.872 | 0.053 |
| | **CaraNet** | **0.773** | **0.689** | 0.729 | **0.853** | **0.902** | **0.042** |
| ETIS | UNet | 0.398 | 0.335 | 0.366 | 0.684 | 0.740 | 0.036 |
| | UNet++ | 0.401 | 0.344 | 0.390 | 0.683 | 0.776 | 0.035 |
| | SFA | 0.297 | 0.217 | 0.231 | 0.557 | 0.633 | 0.109 |
| | PraNet | 0.628 | 0.567 | 0.600 | 0.794 | 0.841 | 0.031 |
| | CCBANet | 0.677 | 0.610 | 0.640 | 0.800 | 0.838 | 0.028 |
| | DS-TransUNet | 0.675 | 0.592 | 0.625 | 0.802 | 0.859 | 0.023 |
| | **CaraNet** | **0.747** | **0.672** | **0.709** | **0.868** | **0.894** | **0.017** |
| CVC-T | UNet | 0.710 | 0.627 | 0.684 | 0.843 | 0.876 | 0.022 |
| | UNet++ | 0.707 | 0.624 | 0.687 | 0.839 | 0.898 | 0.018 |
| | SFA | 0.467 | 0.329 | 0.341 | 0.640 | 0.817 | 0.065 |
| | PraNet | 0.871 | 0.797 | 0.843 | 0.925 | 0.972 | 0.010 |
| | CCBANet | 0.903 | 0.833 | 0.881 | 0.933 | 0.986 | 0.007 |
| | DS-TransUNet | 0.880 | 0.798 | 0.854 | 0.920 | 0.978 | 0.007 |
| | **CaraNet** | **0.903** | **0.838** | **0.887** | **0.940** | **0.989** | **0.007** |

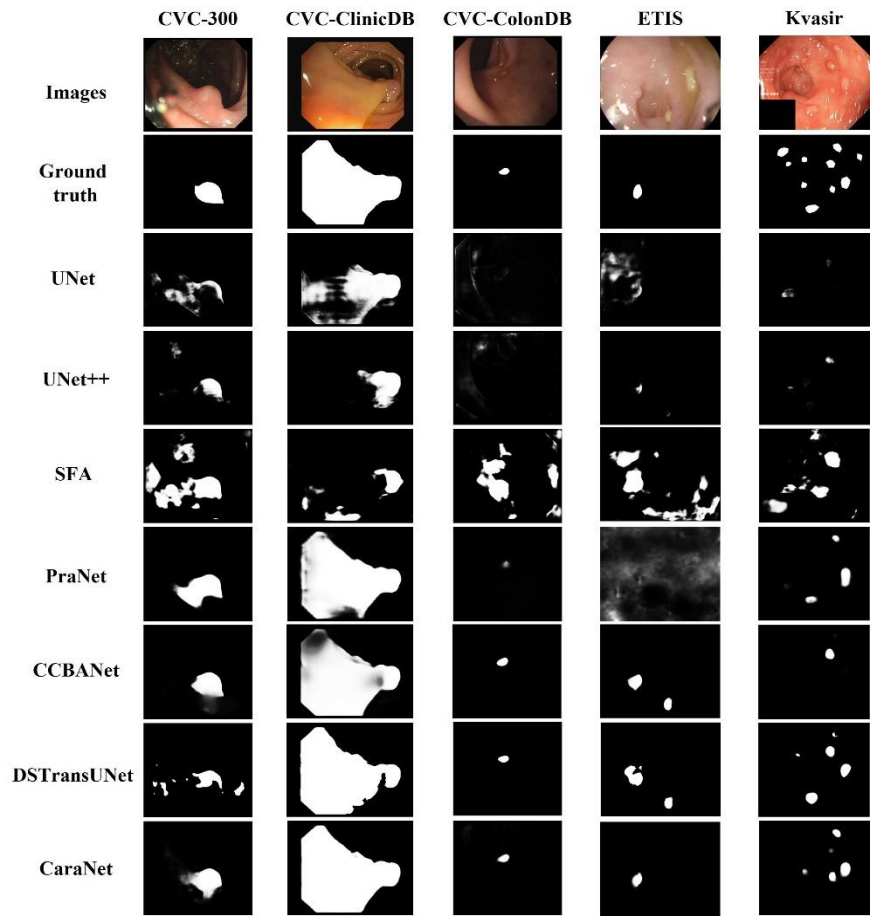

Figure 6. Polyp segmentation results

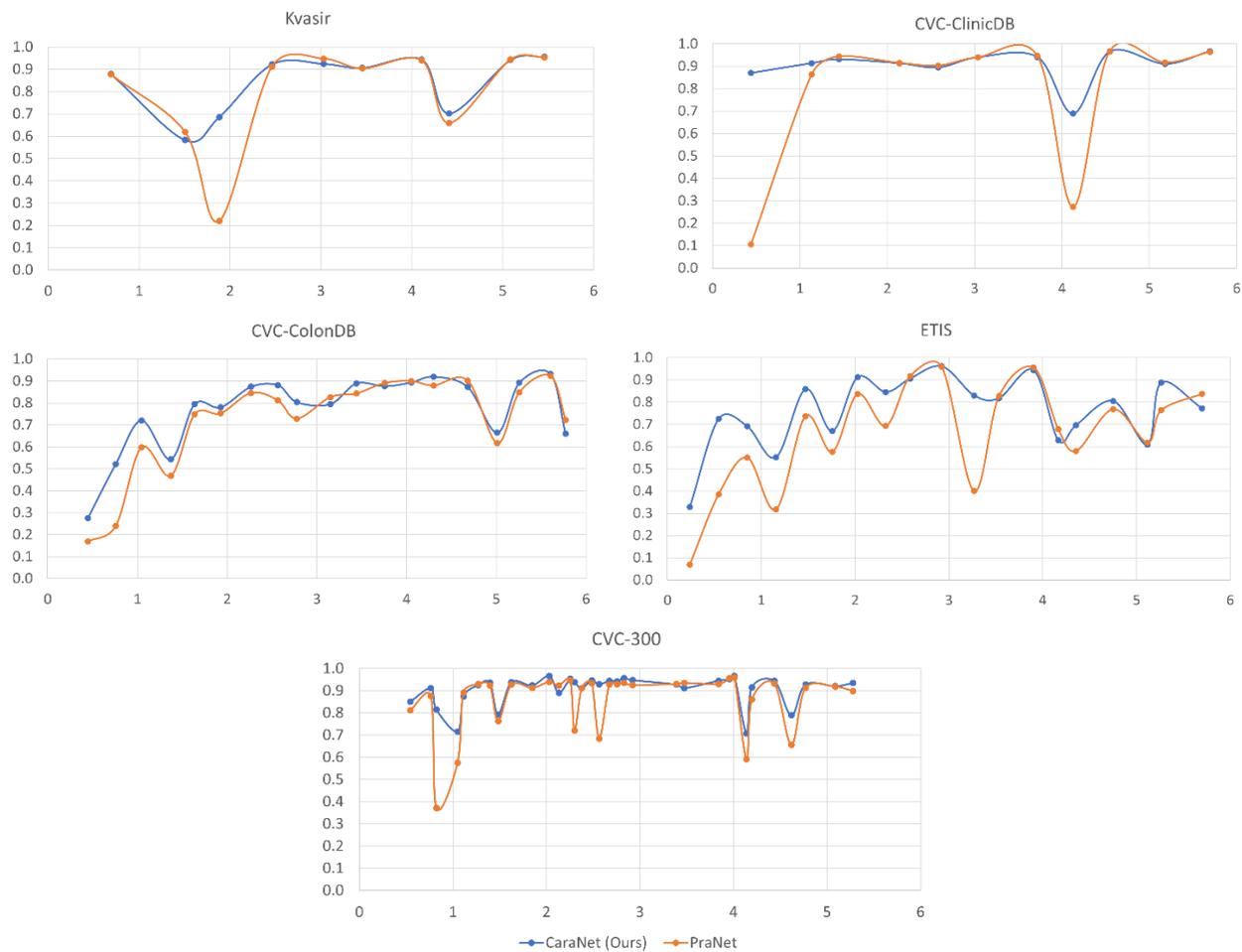

Figure 7. The performance of CaraNet and PraNet for small object segmentation. More discussion about small object segmentation analysis is in Sec. 2.6. For each subfigure, the x-axis is the proportion size (%) of polyp and the y-axis is the averaged mean Dice coefficient. Subfigures show performance vs. size on the five polyp datasets, which are Kvasir, CVC-ClinicDB, CVC-ColonDB, ETIS, and CVC-300. Blue line is for our CaraNet and orange is for the PraNet, showing only the results of small polyp sizes (<6%). We can find that our CaraNet overperforms the PraNet for most cases of small size polyps from the five datasets.

Table 3. Quantitative results on brain tumor (BraTS 2018) dataset.

| Methods | mean Dice | mean IoU | $F_\beta^w$ | $S_\alpha$ | $E_\phi^{max}$ | MAE |
|---|---|---|---|---|---|---|
| CaraNet (Ours) | **0.631** | **0.507** | **0.629** | **0.786** | **0.927** | 0.003 |
| PraNet (MICCAI'20) | 0.619 | 0.494 | 0.606 | 0.776 | 0.920 | 0.003 |

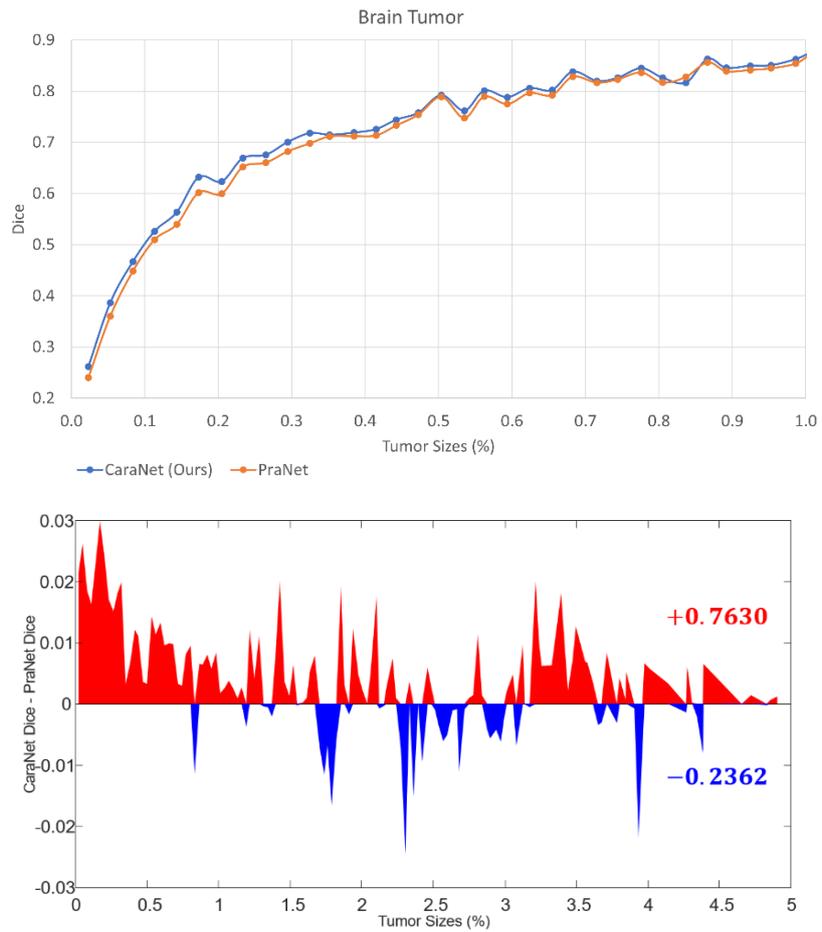

Figure 8. Performance vs. Size on brain tumor datasets. The x-axis is the proportion size (%) of tumor. *Upper figure:* y-axis is the mean Dice coefficient results of our CaraNet and PraNet. For the very small tumor sizes ($\leqslant 1\%$), almost all results of CaraNet are better than PraNet. *Lower figure:* the y-axis is the difference between the averaged mean Dice coefficient results of CaraNet and PraNet. Red indicates the Dice value of CaraNet is greater than that of PraNet; blue shows the opposite.

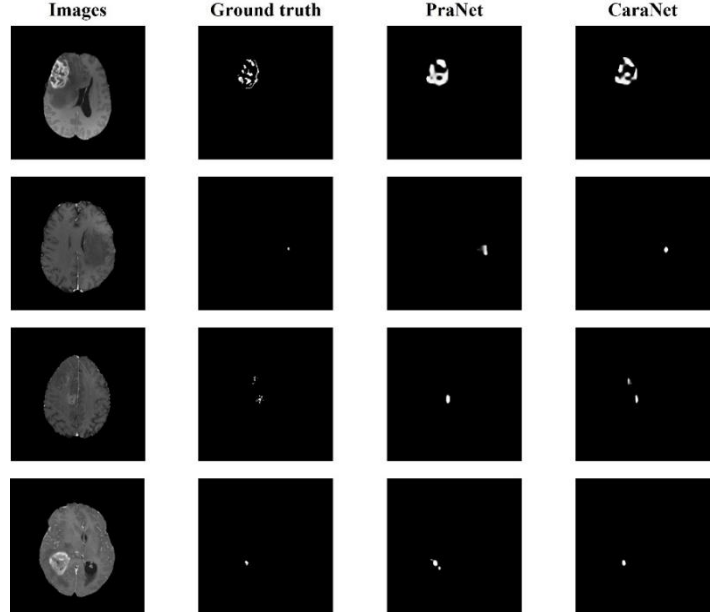

Figure 9. Brain tumor segmentation results

To further evaluate the effectiveness of CaraNet for small-object segmentation, we conducted another experiment using the brain tumor dataset (BraTS 2018). The polyp datasets lack extremely small objects (the minimum is about 0.11%) and do not have enough small samples (like Kvasir and CVC-ClinicDB, in Figure 7, there are fewer samples in the range of small sizes). The brain tumor dataset was created from the BraTS 2018 database by slicing 2D images from the "T1ce" source with "NET" type labels. We randomly select 60% of the images as the training set and the remainder as the testing dataset. Altogether, 3231 images with proportions of tumor sizes ranging from 0.01% – 4.91% were in the testing dataset. Table 3, Figure 8, and Figure 9 show the comparison result. We compared CaraNet with only PraNet for the same reason stated above. Clearly, our CaraNet performed better, especially for the extremely small cases (range 0.01% – 0.1% in Figure 8 and the red area indicates that the Dice value of CaraNet is greater than that of PraNet; blue area shows the opposite. Values on the right show the summations of all red and blue differential values).

### 3.5 Ablation study

Table 4. Quantitative results on Kvasir for different dilation rate

| Dilation Rate | Mean Dice |
|---|---|
| 0 | 0.909 |
| 4 | 0.908 |
| **8** | **0.918** |
| 16 | 0.913 |
| 32 | 0.907 |

To search the best dilation rate for our CFP module, we set up experiments and choose different dilation rates from 0 to 32. The performance of CaraNet with different dilation rate are testing on Kvasir testing set as shown in Table 4. When we choose small dilation rates like 0 and 4, the dilations rates for each channel are $\{r_1, r_2, r_3, r_4\} = \{0,0,0,0\}$ and $\{r_1, r_2, r_3, r_4\} = \{0,1,2,4\}$. The CFP module focuses on local information but ignore the global one, thus the accuracy is about 1% lower than the best one. When large dilation rates are chosen like 16 and 32, the dilation rates for each channel are $\{r_1, r_2, r_3, r_4\} = \{0,4,8,16\}$ and $\{r_1, r_2, r_3, r_4\} = \{0,8,16,32\}$. There only one channel focus on local features which is unreasonable for small medical object detection. When we choose a fairish dilation rate like 8 which $\{r_1, r_2, r_3, r_4\} = \{0,2,4,8\}$, the weights of local and global information can be balanced and then achieves best performance.

We further conduct ablation studies to demonstrate the effectiveness of our proposed CFP module and Axial Reverse Attention (A-RA) module five public endoscopy segmentation datasets, and we choose same six measurement metrics as in Section 3.4 for evaluation.

We first conduct an experiment to evaluate CaraNet without CFP module. As shown in Table 5, the performance without CFP module drops sharply on five public datasets. In particular, the mDice achieves about 8% reduction on ETIS dataset, which strongly verifies that CFP module can effectively detect local-to-global feature due to its various sizes receptive field. Then, we evaluate the CaraNet without both CFP module and A-RA module. The mDice continued to decrease about 2%-3%, indicating that our A-RA module enables our CaraNet to accurately distinguish true polyp tissues.

Table 5. Ablation study for CaraNet on Kvasir, CVC-ClinicDB, CVC-ColonDB, ETIS, and CVC-T (test dataset of EndoScene). Note: mDice: mean Dice, mIoU: mean IoU, $F_\beta^w$: weighted dice, $S_\alpha$: structural measurement [46], $E_\phi^{max}$: enhanced alignment [45] and MAE: mean absolute error. ↑ denotes higher the better and ↓ denotes lower the better.

| Dataset | CFP | A-RA | mDice ↑ | mIoU ↑ | $F_\beta^w$ ↑ | $S_\alpha$ ↑ | $E_\phi^{max}$ ↑ | MAE ↓ |
|---|---|---|---|---|---|---|---|---|
| Kvasir | ✘ | ✘ | 0.870 | 0.798 | 0.836 | 0.899 | 0.938 | 0.043 |
|  | ✘ | ✓ | 0.888 | 0.830 | 0.878 | 0.911 | 0.940 | 0.032 |
|  | ✓ | ✓ | **0.918** | **0.865** | **0.909** | **0.929** | **0.968** | **0.023** |
| CVC-ClinicDB | ✘ | ✘ | 0.862 | 0.789 | 0.841 | 0.908 | 0.949 | 0.021 |
|  | ✘ | ✓ | 0.887 | 0.830 | 0.874 | 0.928 | 0.966 | 0.012 |
|  | ✓ | ✓ | **0.936** | **0.887** | **0.931** | **0.954** | **0.991** | **0.007** |
| ColonDB | ✘ | ✘ | 0.681 | 0.586 | 0.595 | 0.797 | 0.865 | 0.063 |
|  | ✘ | ✓ | 0.707 | 0.636 | 0.689 | 0.815 | 0.867 | 0.048 |
|  | ✓ | ✓ | **0.773** | **0.689** | **0.729** | **0.853** | **0.902** | **0.042** |
| ETIS | ✘ | ✘ | 0.646 | 0.570 | 0.587 | 0.790 | 0.833 | 0.055 |
|  | ✘ | ✓ | 0.662 | 0.580 | 0.604 | 0.805 | 0.869 | 0.032 |
|  | ✓ | ✓ | **0.747** | **0.672** | **0.709** | **0.868** | **0.894** | **0.017** |
| CVC-T | ✘ | ✘ | 0.839 | 0.765 | 0.802 | 0.909 | 0.943 | 0.022 |
|  | ✘ | ✓ | 0.870 | 0.803 | 0.839 | 0.917 | 0.965 | 0.009 |
|  | ✓ | ✓ | **0.903** | **0.838** | **0.887** | **0.940** | **0.989** | **0.007** |

## 4. DISCUSSION

We propose a novel deep-learning based segmentation model – CaraNet, by combining the Axial Reverse Attention and Channel-wise Feature Pyramid (CFP) modules. This new method can help improve the performance of the segmentation of small medical objects. Through the experiments, we show that CaraNet outperforms the most famous models by a large margin overall for six measurement metrics. As shown by the polyp segmentation results, CaraNet not only produces high quality segmentation on samples of large polyps, but also performs well for small and multi small-object segmentation. Figure 9 shows some results of extremely small tumor segmentation from the BraTS 2018 dataset. The advantage of the CaraNet in segmenting small single- and multi-objects is evident. In addition, compared with the recent state-of-the-art network, PraNet, CaraNet provides a more precise prediction for the most-challenging cases.

We also introduce the process to evaluate segmentation models according to the size of objects. We consider the object's size using the size ratio including the sizes of objects and the whole image. In this study, we assume the size ratios of "small objects" are less than 5%. However, the definition of "small objects" is not specified. Since few studies have fully considered the sizes of objects and the small-object problems in medical imaging, we could further study this question in future work. Preliminarily, the small size (watershed) could be defined by the Performance vs. Size curves. Figure 10 shows an example on ETIS datasets. The watershed of small size could be defined at 9.8%. After that point, the performance generally increase/change slower and is more stable. The definition of small area may depend on datasets, segmentation models, and object shapes; but if these conditions are fixed, it is feasible to make fair comparisons. The definition of small area discussed here may not be perfect, but it is worth paying attention to the model's performance on small size cases besides the whole dataset. Figure 10 also indicates that the overall performance of segmentation depends on size ranges. If we remove the small size cases, the overall performance will be greatly improved.

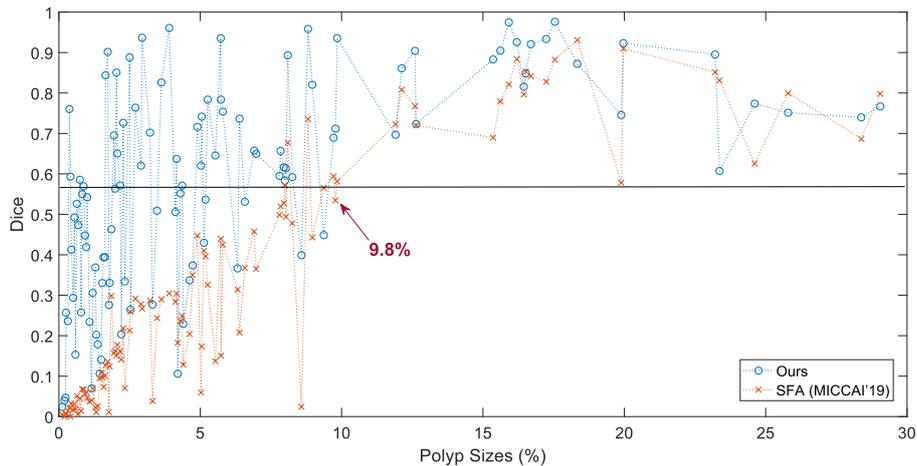

Figure 10. Performance vs. Size on ETIS datasets. The x-axis is the proportion size (%) of polyp; y-axis is the mean Dice coefficient. Blue is for our CaraNet and orange is for the SFA model. Unlike Figure 7, it shows the results of all sizes.

Although CaraNet achieves good improvement on medical image segmentation tasks, there are still some limitations and potentials to optimize the model. For example, using bilinear interpolation to up-sample feature maps cannot avoid the loss of some useful information and lead to a coarse boundary. It can be improved by applying a deconvolutional layer. In addition, the backbone of CaraNet is pre-trained on ImageNet, which contains natural images that are very different from medical images. Moreover, the sliced brain MRI data also cause loss of spatial information between the voxels; that may influence the accuracy of small tumor detection. In our future work, we will use the Model Genesis [47] as a 3-D backbone to replace the Res2Net (2-D backbone) and adjust the CaraNet to employ it on 3-D medical imaging segmentation to build a 3-D version CaraNet for more accurate CT or MRI image segmentation. Segmentation of 3-D medical images is of growing interest; we believe CaraNet will address that problem successfully.

## 5. CONCLUSION

We have proposed a novel neural network, CaraNet, for small medical object segmentation. We use a partial decoder to roughly localize polyp position, improve the adaptability of the CaraNet to detect different sizes objects by using a CFP module and finally refine the polyp segmentation by A-RA module. From the overall segmentation accuracy, we can find that CaraNet outperforms all state-of-the-art approaches by at least 2% (mean dice accuracy). For the early diagnosis dataset (ETIS) which contains many small polyps, however, CaraNet can reach 74.7% mean dice accuracy, which is about 12% higher than PraNet. For the extremely small object segmentation dataset (BraTS 2018), CaraNet can achieve 3% higher than PraNet. When evaluating segmentation models according to the size of objects, CaraNet outperforms PraNet for small objects in all six datasets we used. In addition to the average accuracy through the whole dataset, to evaluate/compare segmentation models according to sizes of objects could show more characteristics of models; especially, their performance on small areas. And this method can find or verify models that are good for small objects segmentation.

## DISCLOSURES

No conflicts of interest.

## References

[1] Ronneberger, O., Fischer, P., & Brox, T. (2015, October). U-net: Convolutional networks for biomedical image segmentation. In International Conference on Medical image computing and computer-assisted intervention (pp. 234-241). Springer, Cham.
[2] Zhou, Z., Siddiquee, M. M. R., Tajbakhsh, N., & Liang, J. (2018). Unet++: A nested u-net architecture for medical image segmentation. In Deep learning in medical image analysis and multimodal learning for clinical decision support (pp. 3-11). Springer, Cham.

[3] Jha, D., Smedsrud, P. H., Riegler, M. A., Johansen, D., De Lange, T., Halvorsen, P., & Johansen, H. D. (2019, December). Resunet++: An advanced architecture for medical image segmentation. In 2019 IEEE International Symposium on Multimedia (ISM) (pp. 225-2255). IEEE.

[4] Fan, D. P., Ji, G. P., Zhou, T., Chen, G., Fu, H., Shen, J., & Shao, L. (2020, October). Pranet: Parallel reverse attention network for polyp segmentation. In International Conference on Medical Image Computing and Computer-Assisted Intervention (pp. 263-273). Springer, Cham.

[5] Menze, B. H., Jakab, A., Bauer, S., Kalpathy-Cramer, J., Farahani, K., Kirby, J., ... & Van Leemput, K. (2014). The multimodal brain tumor image segmentation benchmark (BRATS). IEEE transactions on medical imaging, 34(10), 1993-2024.

[6] Bakas, S., Akbari, H., Sotiras, A., Bilello, M., Rozycki, M., Kirby, J. S., ... & Davatzikos, C. (2017). Advancing the cancer genome atlas glioma MRI collections with expert segmentation labels and radiomic features. Scientific data, 4(1), 1-13.

[7] Bakas, S., Reyes, M., Jakab, A., Bauer, S., Rempfler, M., Crimi, A., ... & Jambawalikar, S. R. (2018). Identifying the best machine learning algorithms for brain tumor segmentation, progression assessment, and overall survival prediction in the BRATS challenge. arXiv preprint arXiv:1811.02629.

[8] Heller, N., Sathianathen, N., Kalapara, A., Walczak, E., Moore, K., Kaluzniak, H., ... & Weight, C. (2019). The kits19 challenge data: 300 kidney tumor cases with clinical context, ct semantic segmentations, and surgical outcomes. arXiv preprint arXiv:1904.00445.

[9] Roth, H., Xu, Z., Diez, C. T., Jacob, R. S., Zember, J., Molto, J., ... & Linguraru, M. (2021). Rapid Artificial Intelligence Solutions in a Pandemic-The COVID-19-20 Lung CT Lesion Segmentation Challenge.

[10] An, P., Xu, S., Harmon, S., Turkbey, E., Sanford, T., Amalou, A., ... & Wood, B. (2020). CT Images in Covid-19 [Data set]. The Cancer Imaging Archive. Mode of access: https://doi. org/10.7937/tcia.

[11] He, K., Zhang, X., Ren, S., & Sun, J. (2016). Deep residual learning for image recognition. In Proceedings of the IEEE conference on computer vision and pattern recognition (pp. 770-778).

[12] Xie, S., Girshick, R., Dollár, P., Tu, Z., & He, K. (2017). Aggregated residual transformations for deep neural networks. In Proceedings of the IEEE conference on computer vision and pattern recognition (pp. 1492-1500).

[13] Huang, G., Liu, Z., Van Der Maaten, L., & Weinberger, K. Q. (2017). Densely connected convolutional networks. In Proceedings of the IEEE conference on computer vision and pattern recognition (pp. 4700-4708).

[14] Szegedy, C., Liu, W., Jia, Y., Sermanet, P., Reed, S., Anguelov, D., ... & Rabinovich, A. (2015). Going deeper with convolutions. In Proceedings of the IEEE conference on computer vision and pattern recognition (pp. 1-9).

[15] Zhao, H., Shi, J., Qi, X., Wang, X., & Jia, J. (2017). Pyramid scene parsing network. In Proceedings of the IEEE conference on computer vision and pattern recognition (pp. 2881-2890).

[16] Chen, L. C., Papandreou, G., Schroff, F., & Adam, H. (2017). Rethinking atrous convolution for semantic image segmentation. arXiv preprint arXiv:1706.05587.

[17] Hesamian, M. H., Jia, W., He, X., & Kennedy, P. (2019). Deep learning techniques for medical image segmentation: achievements and challenges. Journal of digital imaging, 32(4), 582-596.

[18] Ngo, D. K., Tran, M. T., Kim, S. H., Yang, H. J., & Lee, G. S. (2020). Multi-task learning for small brain tumor segmentation from MRI. Applied Sciences, 10(21), 7790.

[19] Taghanaki, S. A., Abhishek, K., Cohen, J. P., Cohen-Adad, J., & Hamarneh, G. (2021). Deep semantic segmentation of natural and medical images: a review. Artificial Intelligence Review, 54(1), 137-178.

[20] Vaswani, A., Shazeer, N., Parmar, N., Uszkoreit, J., Jones, L., Gomez, A. N., ... & Polosukhin, I. (2017). Attention is all you need. In Advances in neural information processing systems (pp. 5998-6008).

[21] Dosovitskiy, A., Beyer, L., Kolesnikov, A., Weissenborn, D., Zhai, X., Unterthiner, T., ... & Houlsby, N. (2020). An image is worth 16x16 words: Transformers for image recognition at scale. arXiv preprint arXiv:2010.11929.

[22] Valanarasu, J. M. J., Oza, P., Hacihaliloglu, I., & Patel, V. M. (2021). Medical transformer: Gated axial-attention for medical image segmentation. arXiv preprint arXiv:2102.10662.

[23] Chen, J., Lu, Y., Yu, Q., Luo, X., Adeli, E., Wang, Y., ... & Zhou, Y. (2021). Transunet: Transformers make strong encoders for medical image segmentation. arXiv preprint arXiv:2102.04306.

[24] Xie, Y., Zhang, J., Shen, C., & Xia, Y. (2021). CoTr: Efficiently Bridging CNN and Transformer for 3D Medical Image Segmentation. arXiv preprint arXiv:2103.03024.

[25] Long, J., Shelhamer, E., & Darrell, T. (2015). Fully convolutional networks for semantic segmentation. In Proceedings of the IEEE conference on computer vision and pattern recognition (pp. 3431-3440).

[26] Fan, D. P., Ji, G. P., Zhou, T., Chen, G., Fu, H., Shen, J., & Shao, L. (2020, October). Pranet: Parallel reverse attention network for polyp segmentation. In International Conference on Medical Image Computing and Computer-Assisted

[27] Dahl, G. E., Yu, D., Deng, L., & Acero, A. (2011). Context-dependent pre-trained deep neural networks for large-vocabulary speech recognition. IEEE Transactions on audio, speech, and language processing, 20(1), 30-42.

[28] Deng, J., Dong, W., Socher, R., Li, L. J., Li, K., & Fei-Fei, L. (2020). Imagenet: A large-scale hierarchical image database, 2009. In IEEE Conference on Computer Vision and Pattern Recognition (CVPR) (pp. 248-255).

[29] Gao, S., Cheng, M. M., Zhao, K., Zhang, X. Y., Yang, M. H., & Torr, P. H. (2019). Res2net: A new multi-scale backbone architecture. IEEE transactions on pattern analysis and machine intelligence.

[30] Wu, Z., Su, L., & Huang, Q. (2019). Cascaded partial decoder for fast and accurate salient object detection. In Proceedings of the IEEE/CVF Conference on Computer Vision and Pattern Recognition (pp. 3907-3916).

[31] Zhao, H., Shi, J., Qi, X., Wang, X., & Jia, J. (2017). Pyramid scene parsing network. In Proceedings of the IEEE conference on computer vision and pattern recognition (pp. 2881-2890).

[32] Lin, T. Y., Dollár, P., Girshick, R., He, K., Hariharan, B., & Belongie, S. (2017). Feature pyramid networks for object detection. In Proceedings of the IEEE conference on computer vision and pattern recognition (pp. 2117-2125).

[33] Lou, A., & Loew, M. (2021). CFPNet: Channel-wise Feature Pyramid for Real-Time Semantic Segmentation. arXiv preprint arXiv:2103.12212.

[34] Lou, A., Guan, S., & Loew, M. (2021). CFPNet-M: A Light-Weight Encoder-Decoder Based Network for Multimodal Biomedical Image Real-Time Segmentation. arXiv preprint arXiv:2105.04075.

[35] Mehta, S., Rastegari, M., Caspi, A., Shapiro, L., & Hajishirzi, H. (2018). Espnet: Efficient spatial pyramid of dilated convolutions for semantic segmentation. In Proceedings of the european conference on computer vision (ECCV) (pp. 552-568).

[36] Wang, H., Zhu, Y., Green, B., Adam, H., Yuille, A., & Chen, L. C. (2020, August). Axial-deeplab: Stand-alone axial-attention for panoptic segmentation. In European Conference on Computer Vision (pp. 108-126). Springer, Cham.

[37] Chen, S., Tan, X., Wang, B., & Hu, X. (2018). Reverse attention for salient object detection. In Proceedings of the European Conference on Computer Vision (ECCV) (pp. 234-250).

[38] Silva, J., Histace, A., Romain, O., Dray, X., & Granado, B. (2014). Toward embedded detection of polyps in wce images for early diagnosis of colorectal cancer. International journal of computer assisted radiology and surgery, 9(2), 283-293.

[39] Bernal, J., Sánchez, F. J., Fernández-Esparrach, G., Gil, D., Rodríguez, C., & Vilariño, F. (2015). WM-DOVA maps for accurate polyp highlighting in colonoscopy: Validation vs. saliency maps from physicians. Computerized Medical Imaging and Graphics, 43, 99-111.

[40] Tajbakhsh, N., Gurudu, S. R., & Liang, J. (2015). Automated polyp detection in colonoscopy videos using shape and context information. IEEE transactions on medical imaging, 35(2), 630-644.

[41] Vázquez, D., Bernal, J., Sánchez, F. J., Fernández-Esparrach, G., López, A. M., Romero, A., ... & Courville, A. (2017). A benchmark for endoluminal scene segmentation of colonoscopy images. Journal of healthcare engineering, 2017.

[42] Jha, D., Smedsrud, P. H., Riegler, M. A., Halvorsen, P., de Lange, T., Johansen, D., & Johansen, H. D. (2020, January). Kvasir-seg: A segmented polyp dataset. In International Conference on Multimedia Modeling (pp. 451-462). Springer, Cham.

[43] Zhang, Z., Liu, Q., & Wang, Y. (2018). Road extraction by deep residual u-net. IEEE Geoscience and Remote Sensing Letters, 15(5), 749-753.

[44] Fang, Y., Chen, C., Yuan, Y., & Tong, K. Y. (2019, October). Selective feature aggregation network with area-boundary constraints for polyp segmentation. In International Conference on Medical Image Computing and Computer-Assisted Intervention (pp. 302-310). Springer, Cham.

[45] Fan, D. P., Gong, C., Cao, Y., Ren, B., Cheng, M. M., & Borji, A. (2018). Enhanced-alignment measure for binary foreground map evaluation. arXiv preprint arXiv:1805.10421.

[46] Fan, D. P., Cheng, M. M., Liu, Y., Li, T., & Borji, A. (2017). Structure-measure: A new way to evaluate foreground maps. In Proceedings of the IEEE international conference on computer vision (pp. 4548-4557).

[47] Zhou, Z., Sodha, V., Pang, J., Gotway, M. B., & Liang, J. (2021). Models genesis. Medical image analysis, 67, 101840.

[48] B. H. Menze, A. Jakab, S. Bauer, J. Kalpathy-Cramer, K. Farahani, J. Kirby, et al. (2015). "The Multimodal Brain Tumor Image Segmentation Benchmark (BRATS)", IEEE Transactions on Medical Imaging 34(10), 1993-2024 DOI: 10.1109/TMI.2014.2377694

[49] S. Bakas, H. Akbari, A. Sotiras, M. Bilello, M. Rozycki, J.S. Kirby, et al. (2017). "Advancing The Cancer Genome Atlas glioma MRI collections with expert segmentation labels and radiomic features", Nature Scientific Data, 4:170117 DOI: 10.1038/sdata.2017.117

[50] Nguyen, T. C., Nguyen, T. P., Diep, G. H., Tran-Dinh, A. H., Nguyen, T. V., & Tran, M. T. (2021, September). Ccbanet:

Cascading context and balancing attention for polyp segmentation. In *International Conference on Medical Image Computing and Computer-Assisted Intervention* (pp. 633-643). Springer, Cham.

[51] Lin, A., Chen, B., Xu, J., Zhang, Z., Lu, G., & Zhang, D. (2022). Ds-transunet: Dual swin transformer u-net for medical image segmentation. *IEEE Transactions on Instrumentation and Measurement*.